\documentclass[floatfix,notitlepage,aps,prd,twocolumn,superscriptaddress,altaffilletter]{revtex4-1}

\usepackage{amsmath}
\usepackage{tensor}
\usepackage{graphicx}
\usepackage{dcolumn}
\usepackage{bm}
\usepackage{hyperref}
\usepackage{xcolor}
\usepackage{subfigure}
\usepackage{amsmath}
\usepackage{amssymb}
\usepackage{lipsum}

\newcommand{\ddo}{D$_2$O}

\begin{document}
\preprint{APS/123-QED}

\title{Search for \textit{hep} solar neutrinos and the diffuse
supernova neutrino background using all three phases of the
Sudbury Neutrino Observatory}

%bibnotes puts the present addresses in with the references.

%Authorlist v.20.2 hep April 2020.
\newcommand{\dec}{Deceased}
\newcommand{\alta}{Department of Physics, University of 
Alberta, Edmonton, Alberta, T6G 2R3, Canada}
\newcommand{\chicago}{Department of Physics, University of 
Chicago, Chicago IL 60637}
\newcommand{\ubc}{Department of Physics and Astronomy, University of 
British Columbia, Vancouver, BC V6T 1Z1, Canada}
\newcommand{\bnl}{Chemistry Department, Brookhaven National 
Laboratory,  Upton, NY 11973-5000}
\newcommand{\carleton}{Ottawa-Carleton Institute for Physics, Department of Physics, Carleton University, Ottawa, Ontario K1S 5B6, Canada}
\newcommand{\carletona}{Department of Physics, Carleton University, Ottawa, Ontario, Canada}
\newcommand{\uog}{Physics Department, University of Guelph,  
Guelph, Ontario N1G 2W1, Canada}
\newcommand{\lu}{Department of Physics and Astronomy, Laurentian 
University, Sudbury, Ontario P3E 2C6, Canada}
\newcommand{\lbnl}{Institute for Nuclear and Particle Astrophysics and 
Nuclear Science Division, Lawrence Berkeley National Laboratory, Berkeley, CA 94720-8153}
\newcommand{\lbla}{ Lawrence Berkeley National Laboratory, Berkeley, CA}
\newcommand{\lanl}{Los Alamos National Laboratory, Los Alamos, NM 87545}
\newcommand{\llnl}{Lawrence Livermore National Laboratory, Livermore, CA}
\newcommand{\lanla}{Los Alamos National Laboratory, Los Alamos, NM 87545}
\newcommand{\oxford}{Department of Physics, University of Oxford, 
Denys Wilkinson Building, Keble Road, Oxford OX1 3RH, UK}
\newcommand{\penn}{Department of Physics and Astronomy, University of 
Pennsylvania, Philadelphia, PA 19104-6396}
\newcommand{\pennx}{Department of Physics and Astronomy, University of 
Pennsylvania, Philadelphia, PA}
\newcommand{\queens}{Department of Physics, Queen's University, 
Kingston, Ontario K7L 3N6, Canada}
\newcommand{\uw}{Center for Experimental Nuclear Physics and Astrophysics, 
and Department of Physics, University of Washington, Seattle, WA 98195}
\newcommand{\uwx}{Center for Experimental Nuclear Physics and Astrophysics, 
and Department of Physics, University of Washington, Seattle, WA}
\newcommand{\uta}{Department of Physics, University of Texas at Austin, Austin, TX 78712-0264}
\newcommand{\triumf}{TRIUMF, 4004 Wesbrook Mall, Vancouver, BC V6T 2A3, Canada}
\newcommand{\ralimp}{Rutherford Appleton Laboratory, Chilton, Didcot, UK} %OX11 0QX
\newcommand{\iusb}{Department of Physics and Astronomy, Indiana University, South Bend, IN}
\newcommand{\fnal}{Fermilab, Batavia, IL}
\newcommand{\uo}{Department of Physics and Astronomy, University of Oregon, Eugene, OR}
\newcommand{\hu}{Department of Physics, Hiroshima University, Hiroshima, Japan}
\newcommand{\slac}{Stanford Linear Accelerator Center, Menlo Park, CA}
\newcommand{\mac}{Department of Physics, McMaster University, Hamilton, ON}
\newcommand{\doe}{US Department of Energy, Germantown, MD}
\newcommand{\lund}{Department of Physics, Lund University, Lund, Sweden}
\newcommand{\mpi}{Max-Planck-Institut for Nuclear Physics, Heidelberg, Germany}
\newcommand{\uom}{Ren\'{e} J.A. L\'{e}vesque Laboratory, Universit\'{e} de Montr\'{e}al, Montreal, PQ}
\newcommand{\cwru}{Department of Physics, Case Western Reserve University, Cleveland, OH}
\newcommand{\pnnl}{Pacific Northwest National Laboratory, Richland, WA}
\newcommand{\uc}{Department of Physics, University of Chicago, Chicago, IL}
\newcommand{\mitt}{Laboratory for Nuclear Science, Massachusetts Institute of Technology, Cambridge, MA 02139}
\newcommand{\ucsd}{Department of Physics, University of California at San Diego, La Jolla, CA }
\newcommand{	\lsu	}{Department of Physics and Astronomy, Louisiana State University, Baton Rouge, LA 70803}
\newcommand{\imp}{Imperial College, London, UK}%SW7 2AZ
\newcommand{\uci}{Department of Physics, University of California, Irvine, CA 92717}
\newcommand{\ucia}{Department of Physics, University of California, Irvine, CA}
\newcommand{\suss}{Department of Physics and Astronomy, University of Sussex, Brighton  BN1 9QH, UK}
\newcommand{\sussx}{Department of Physics and Astronomy, University of Sussex, Brighton, UK}
\newcommand{\lifep}{Laborat\'{o}rio de Instrumenta\c{c}\~{a}o e F\'{\i}sica Experimental de
Part\'{\i}culas, Avenida Elias Garcia 14, 1$^{\circ}$, 1000-149 Lisboa, Portugal}
\newcommand{\lipx}{Laborat\'{o}rio de Instrumenta\c{c}\~{a}o e F\'{\i}sica Experimental de
Part\'{\i}culas,  Lisboa, Portugal}
\newcommand{\hku}{Department of Physics, The University of Hong Kong, Hong Kong.}
\newcommand{\aecl}{Atomic Energy of Canada, Limited, Chalk River Laboratories, Chalk River, ON K0J 1J0, Canada}
\newcommand{\nrc}{National Research Council of Canada, Ottawa, ON K1A 0R6, Canada}
\newcommand{\princeton}{Department of Physics, Princeton University, Princeton, NJ 08544}
\newcommand{\birkbeck}{Birkbeck College, University of London, Malet Road, London WC1E 7HX, UK}
\newcommand{\snoi}{SNOLAB, Lively, ON P3Y 1N2, Canada}
\newcommand{\snoix}{SNOLAB, Lively,  ON, Canada}
\newcommand{\uba}{University of Buenos Aires, Argentina}
\newcommand{\hvd}{Department of Physics, Harvard University, Cambridge, MA}
\newcommand{\pny}{Goldman Sachs, 85 Broad Street, New York, NY}
\newcommand{\pnv}{Remote Sensing Lab, PO Box 98521, Las Vegas, NV 89193}
\newcommand{\nts}{Nevada National Security Site, Las Vegas, NV}
\newcommand{\psis}{Paul Schiffer Institute, Villigen, Switzerland}
\newcommand{\liverpool}{Department of Physics, University of Liverpool, Liverpool, UK}
\newcommand{\uto}{Department of Physics, University of Toronto, Toronto, ON, Canada}
\newcommand{\uwisc}{Department of Physics, University of Wisconsin, Madison, WI}
\newcommand{\psu}{Department of Physics, Pennsylvania State University,
     University Park, PA}
\newcommand{\anl}{Deparment of Mathematics and Computer Science, Argonne
     National Laboratory, Lemont, IL}
\newcommand{\cornell}{Department of Physics, Cornell University, Ithaca, NY}
\newcommand{\tufts}{Department of Physics and Astronomy, Tufts University, Medford, MA}
\newcommand{\ucd}{Department of Physics, University of California, Davis, CA}
\newcommand{\unc}{Department of Physics, University of North Carolina, Chapel Hill, NC}
\newcommand{\dresden}{Institut f\"{u}r Kern- und Teilchenphysik, Technische Universit\"{a}t Dresden,  Dresden, Germany} %01069 
\newcommand{\isu}{Department of Physics, Idaho State University, Pocatello, ID}
\newcommand{\qmul}{School of Physics and Astronomy, Queen Mary University of London, UK}
\newcommand{\ucsb}{Dept. of Physics, University of California, Santa Barbara, CA}
\newcommand{\cern}{CERN, Geneva, Switzerland}
\newcommand{\utah}{Dept. of Physics, University of Utah, Salt Lake City, UT}
\newcommand{\casa}{Center for Astrophysics and Space Astronomy, University
of Colorado, Boulder, CO}%80309
\newcommand{\susel}{Sanford Underground Research Laboratory, Lead, SD}  %57754
\newcommand{\ntu}{Center of Cosmology and Particle Astrophysics, National Taiwan University, Taiwan}
\newcommand{\berlin}{Institute for Space Sciences, Freie Universit\"{a}t Berlin,
Leibniz-Institute of Freshwater Ecology and Inland Fisheries, Germany}
\newcommand{\bhsu}{Black Hills State University, Spearfish, SD} %57799-9003
\newcommand{\queensa}{Dept.\,of Physics, Queen's University, 
Kingston, Ontario, Canada} % K7L 3N6
\newcommand{\aasu}{Dept.\,of Chemistry and Physics, Armstrong  State University, Savannah, GA}
\newcommand{\ucb}{Physics Department, University of California at Berkeley, Berkeley, CA 94720-7300}
\newcommand{\ucbx}{Physics Department, University of California at Berkeley, and Lawrence Berkeley National Laboratory, Berkeley, CA}
\newcommand{\mcgill}{Physics Department, McGill University, Montreal, QC, Canada}% H3A 2T8
\newcommand{\columbia}{Columbia University, New York, NY}%10027
\newcommand{\rhul}{Dept. of Physics, Royal Holloway University of London, Egham, Surrey, UK}%  TW20 0EX
\newcommand{\ubama}{Department of Physics and Astronomy, University of Alabama, Tuscaloosa, AL}
\newcommand{\kit}{Institut f\"{u}r Experimentelle Kernphysik, Karlsruher Institut f\"{u}r Technologie, Karlsruhe, Germany}
\newcommand{\winnipeg}{Department of Physics, University of Winnipeg, Winnipeg, Manitoba, Canada}% 515 Portage Ave. R3B 2E9
\newcommand{\kwantlen}{Kwantlen Polytechnic University, Surrey, BC, Canada}% 12666 72nd Ave. V3W 2M8
\newcommand{\sunysb}{Laufer Center, Stony Brook University, Stony Brook, NY}% 11794
\newcommand{\rock}{Rock Creek Group, Washington, DC}
\newcommand{\rcnp}{Research Center for Nuclear Physics, Osaka, Japan}
\newcommand{\usd}{University of South Dakota, Vermillion, SD}
\newcommand{\lancaster}{Physics Department, Lancaster University, Lancaster, UK}%LA1 4YB,
\newcommand{\potsdam}{GFZ German Research Centre for Geosciences, Potsdam, Germany}%14473
\newcommand{\kirchhoff}{Ruprecht-Karls-Universit\"{a}t Heidelberg, Im Neuenheimer Feld 227, Heidelberg, Germany}%*D-69120
\newcommand{\continuum}{Continuum Analytics,  Austin, TX}% 221 W. 6th St. #1550,   78701
\newcommand{\gsu}{Dept. of Physics, Georgia Southern University, Statesboro, GA}
\newcommand{\pelmorex}{Pelmorex Corp., Oakville, ON} %2655 Bristol Circle, Oakville ON L6H7W1
\newcommand{\usaid}{Global Development Lab, U.S. Agency for International Development, Washington DC}%1300 Pennsylvania Ave NW, Washington, DC, 20004.
\newcommand{\bnc}{National Bank of Canada, Montreal, QC, Canada}
\newcommand{\kings}{Department of Physics, King's College London, London, UK} % WC2R 2LS
\newcommand{\uky}{Department of Physics and Astronomy, University of Kentucky, Lexington KY}
\newcommand{\rutgers}{Department of Physics and Astronomy, Rutgers University, Piscataway, NJ}

%%%%%%%%%%%%

%\affiliation{\aecl}
\affiliation{\alta}
\affiliation{\ucb}
\affiliation{\ubc}
\affiliation{\bnl}
%\affiliation{\uci}
\affiliation{\carleton}
\affiliation{\chicago}
\affiliation{\uog}
\affiliation{\lu}
\affiliation{\lbnl}
\affiliation{\lifep}
\affiliation{\lanl}
\affiliation{\lsu}
\affiliation{\mitt}
%\affiliation{\nrc}
\affiliation{\oxford}
\affiliation{\penn}
%\affiliation{\princeton}
\affiliation{\queens}
%\affiliation{\ralimp}
\affiliation{\snoi}
\affiliation{\uta}
\affiliation{\triumf}
\affiliation{\uw}
\author{B.~Aharmim}\affiliation{\lu}
\author{S.\,N.~Ahmed}\affiliation{\queens}
\author{A.\,E.~Anthony}\altaffiliation{Present address: \usaid}\affiliation{\uta}
\author{N.~Barros}\affiliation{\penn}\affiliation{\lifep}
\author{E.\,W.~Beier}\affiliation{\penn}
\author{A.~Bellerive}\affiliation{\carleton}
\author{B.~Beltran}\affiliation{\alta}
\author{M.~Bergevin}\altaffiliation{Present address: \llnl}\affiliation{\lbnl}\affiliation{\uog}
\author{S.\,D.~Biller}\affiliation{\oxford}
\author{E.~Blucher}\affiliation{\chicago}
% ...
			
\collaboration{SNO Collaboration}
\noaffiliation

\date{\today}

\begin{abstract}
A search has been performed for neutrinos from two sources, the $hep$
reaction in the solar $pp$ fusion chain and the $\nu_e$ component of the
diffuse supernova neutrino background (DSNB), using the full dataset of the
Sudbury Neutrino Observatory with a total exposure of 2.47 kton-years after
fiducialization. The $hep$ search is performed using both a
single-bin counting analysis and a likelihood fit. We find a best-fit
flux that is compatible with solar model predictions while remaining
consistent with zero flux, and set a one-sided upper limit of
$\Phi_{hep} < 30\times10^{3}~\mathrm{cm}^{-2}~\mathrm{s}^{-1}$
[90\% credible interval (CI)].
No events are observed in the DSNB search
region, and we set an improved upper bound on the $\nu_e$ component of
the DSNB flux of
$\Phi^\mathrm{DSNB}_{\nu_e} < 19~\textrm{cm}^{-2}~\textrm{s}^{-1}$
(90\% CI) in the energy range $22.9 < E_\nu < 36.9$~MeV.
\end{abstract}

\maketitle

\section{Introduction}
\label{sec:intro}
Solar neutrinos produced in the $pp$ fusion cycle have been studied
extensively by several experiments
\cite{1998ApJ...496..505C, Altmann:833517,
      1999PhLB..447..127H, 2014PhRvD..89k2007B,
      Aharmim:2013hr, SKIV}.
However, the highest energy branch in this cycle, the $hep$ reaction
[${^{3}\mathrm{He}}(p,e^+\nu_e){^4\mathrm{He}}$],
has yet to be directly detected. With a predicted branching
ratio of $\sim2\times10^{-7}$, the flux expected on Earth in the
BSB05(GS98) solar model is
$(7.93\pm1.23)\times10^{3}~\mathrm{cm}^{-2}~\mathrm{s}^{-1}$
\cite{Bahcall:2006ke,Serenelli:2007zz}.
As the $hep$ reaction has the highest end point energy of all solar
neutrinos, and occurs at a relatively large radius in the Sun,
an observation may provide sensitivity to nonstandard
solar models in addition to completing our picture of
the $pp$ chain neutrino fluxes.

Also expected in the energy range above the end point of the $^8$B
solar neutrino spectrum is the diffuse supernova neutrino background
(DSNB), the isotropic neutrino flux from past core-collapse supernovae
\cite{lunardiniDiffuseSupernovaNeutrinos2016,Beacom:2010kk}.
A measurement of the DSNB would provide new data on supernova dynamics
averaged over these past core-collapse events, which would constrain models
and provide context for nearby core collapse supernova events detectable on
an individual basis, such as SN1987A
\cite{Arnett:1989ka,Hirata:1987hu,Bionta:1987qt,Alekseev:1987ej}.
In particular, the total flux provides a measure of the average supernova
luminosity in neutrinos, and the spectrum is dependent on the temperature
at the surface of last scattering. The DSNB signal remains undetected,
and the Sudbury Neutrino Observatory (SNO)
experiment provides unique
sensitivity to the $\nu_e$ component of the flux \cite{2006PhRvC..73c5807B}.

A previous search for the $hep$ and DSNB neutrinos with the SNO
detector used data from the first operating phase, 306.4 live days
with a heavy water (D$_2$O) target \cite{Collaboration:2006wh}.
The present work extends
that counting analysis to the full SNO dataset across all
operating phases, and additionally a spectral fit is performed.
Section \ref{sec:detector} briefly introduces the SNO detector.
Next, Sec. \ref{sec:analysis} describes the dataset, event
selection, and the counting and fit-based analysis methods.
Finally, results are presented in Sec. \ref{sec:results}.

\section{The SNO Detector}
\label{sec:detector}
\begin{figure}
\centering
\includegraphics[width=0.49\textwidth]{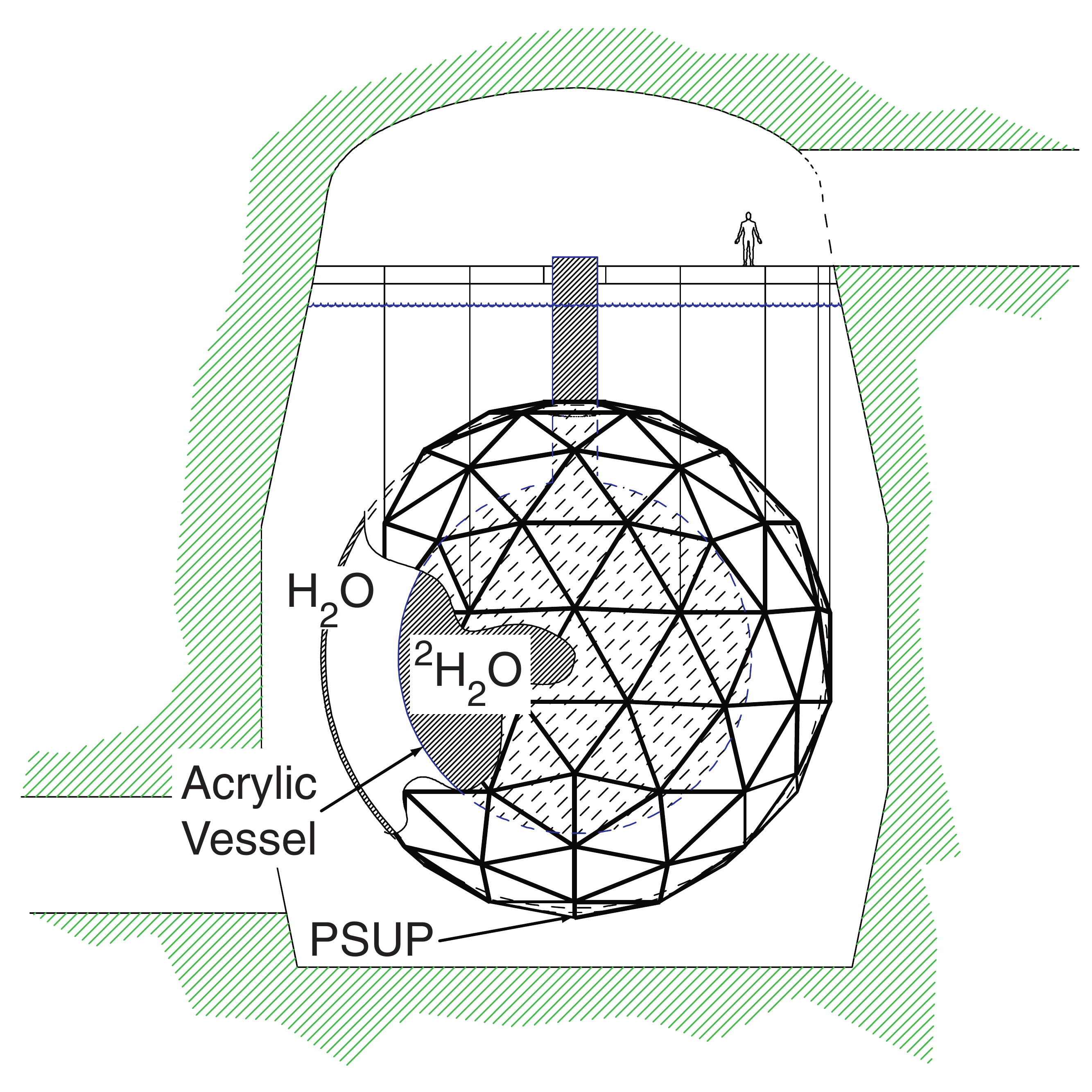}
\caption{The SNO detector \cite{Aharmim:2013hr}.}
\label{fig:detector}
\end{figure}

The SNO detector \cite{Boger:2000vy} consisted of a target volume
enclosed within a transparent acrylic sphere 6 m in radius,
viewed by 9456 inward-looking 8-inch photomultiplier tubes (PMTs) at a
radius of 8.4 m, as illustrated in Fig. \ref{fig:detector}.
The acrylic
vessel and the structure supporting the PMTs (PSUP)
were suspended in a water-filled
cavity, which was additionally instrumented with outward-looking PMTs to
provide an active veto system. In order to shield from cosmic ray
muons and from the
neutrons and radioisotopes resulting from muon interactions, the detector
was located deep underground with a $5890\pm94$ meter water equivalent
rock overburden at the Inco (now Vale) Creighton
mine near Sudbury, Ontario, Canada.

The detector operated in three distinct phases, differing in the
primary mechanism for neutron detection. In the first phase,
the detector was loaded with a very low background heavy
water (\ddo) target. With the \ddo~target,
SNO was sensitive to charged current (CC),
neutral current (NC), and elastic scattering (ES) channels:
\begin{align*}
\nu_e + d \to p + p + e^- - 1.44~\mathrm{MeV} & ~\mathrm{(CC)}, \\
\nu + d   \to p + n + \nu - 2.22~\mathrm{MeV} & ~\mathrm{(NC)}, \\
\nu + e^- \to \nu + e^-                     & ~\mathrm{(ES)}.
\end{align*}
The $hep$ and DSNB searches benefit in particular from the enhancement by a
factor of about 100 of the CC cross section with respect to that for ES, and
from the fact that in the CC interaction, the outgoing electron energy is
strongly correlated with the incoming neutrino energy.

In SNO's second operational phase, the \ddo~was doped with 0.2\% NaCl
by mass, to take advantage of the improved neutron capture cross
section on Cl and
the higher energy and more isotropic deexcitation $\gamma$
cascade. In the third phase, the NaCl was removed and
an array of $^3$He proportional counters (NCDs) was
deployed to further improve neutron detection.
In all three phases, backgrounds due to atmospheric neutrino interactions are
reduced significantly via coincidence tagging of final state neutrons.

\section{Analysis}
\label{sec:analysis}
We performed a single-bin counting analysis in
two different energy ranges, for the $hep$ and DSNB
neutrino signals. Additionally, a maximum likelihood
fit was used to extend the sensitivity of the $hep$
search. The following sections describe the dataset,
event selection criteria, and systematic uncertainties
common to the counting and likelihood analyses, and then
introduce those techniques.

\subsection{Data selection}
This analysis makes use of the entire SNO dataset, across all
three operational phases, with data collected between November
1999 and November 2006. Table \ref{tab:livetime}
indicates the live time for each phase, corresponding to
a total exposure of 2.47 kilotonne-years after fiducialization.
We adopted a pseudo-blind approach in which
the analysis was tuned on Monte Carlo simulations, then
validated on one third of the data randomly sampled in
short blocks of time uniformly distributed throughout the phases.
Finally, with cuts and parameters having been fixed, the full dataset was
reopened for this analysis.

\begin{table}
\begin{ruledtabular}
 \caption{Duration and live time for each operational phase.}
 \label{tab:livetime}
 \centering
 \begin{tabular}{cccc}
  Phase & Target            & Dates & Live time \\
  \hline
  I   & D$_2$O              & 11/1999 -- 5/2001  & 306.4 d \\
  II  & D$_2$O + 0.2\% NaCl & 7/2001 -- 8/2003  & 478.6 d \\
  III & D$_2$O + NCDs       & 11/2004 -- 11/2006 & 387.2 d \\
 \end{tabular}
 \end{ruledtabular}
\end{table}

The set of signal candidate events follows from three
stages of event selection. First, entire runs (approximately
8 hour blocks of live time) are accepted or rejected based on
detector conditions. The same selection is applied as in
Ref. \cite{Ahmad:2002cg} for phase I and Ref.
\cite{Aharmim:2013hr} for phase III. For phase II, the
selection from Ref. \cite{Aharmim:2005hw} is extended to include
periods with higher than average levels of Rn or activated Na, which
presented important backgrounds for the low energy threshold $^8$B
oscillation analyses but are insignificant for the higher-threshold
$hep$ and DSNB searches.

Next, a set of low-level cuts are applied, which
address instrumental background events
as well as coincidences with bursts of events or tagged
muons. The instrumental backgrounds are caused by detector effects,
for example high-voltage discharge of a PMT, or electronic pickup.
Such events tend to have distinct signatures, such as correlations
in the physical locations of electronics channels, which are very
different from signal events. For each phase, the same set of
low-level cuts is used as in previous work \cite{Aharmim:2013hr},
as these have been extensively validated and tuned to optimize
signal efficiency. For this analysis,
signal-like events are further required to be isolated in time:
any candidate event occurring within 250 ms of another candidate event
is rejected. This includes coincidences with any event with a reconstructed
vertex within a 6 m fiducial volume and a kinetic energy above 4~MeV, a
trigger of the external veto, or (in phase III only) a detected signal
in the NCD array.
This reduces background classes that produce coincident electrons, neutrons,
or photons, and in particular targets Michel
electrons following low-energy muons or nuclear deexcitation photons and
atmospheric neutrino CC electrons with neutron followers.

Finally, a series of high level criteria have been developed based on
reconstructed observables, which discriminate the signals of
interest from other physics backgrounds.
The signature of a signal $hep$ or DSNB neutrino interaction is a single
electronlike Cherenkov ring originating within 550~cm of the detector center.
This fiducial volume is chosen to reduce backgrounds associated with
$\gamma$ rays and other
backgrounds due to the materials surrounding the target volume.
Signal Cherenkov rings are highly anisotropic, at a level quantified by the
variable
$\beta_{14}$ previously described in Ref. \cite{Aharmim:2010kt}.
The fraction of PMTs hit within a narrow prompt time window is
calculated as the in-time ratio (ITR). This variable can discriminate
between well-reconstructed single-ring events or multiring events due to
a pileup of interactions or particles. To further discriminate single
electronlike events, three Kolmogorov-Smirnov (KS) tests are used.
The first simply tests the compatibility of the azimuthal distribution of
hits around the reconstructed direction relative to a flat distribution.
The second test is a two-dimensional extension that includes the polar angle
and compares to a probability distribution derived from
calibration data, accounting for energy dependence in the polar angle and
solid angle effects in the azimuthal angle.
A final test compares the time-of-flight
corrected PMT hit times for hits inside the Cherenkov ring to a
template distribution also extracted from calibration data.
Cuts on these parameters have been adjusted relative to previous
SNO analyses as described in Sec. \ref{sec:counting},
as both the energy regime ($>15$~MeV) and the
objectives (rejection of atmospheric neutrino backgrounds)
differ. The distributions in these high-level observables are validated by
comparing simulations to data in the low-energy sideband below the $hep$
region of interest and to calibration data using a signal-like $^{8}$Li
source \cite{Tagg:2002iq}.

\subsection{Monte Carlo simulation}
\label{sec:mc}
The detailed microphysical detector model used in previous SNO
measurements, \textsc{SNOMAN} \cite{Boger:2000vy, Aharmim:2013hr},
was again employed for this analysis.
\textsc{SNOMAN} was used to generate solar neutrino
events, propagate final state particles through the detector
geometry, and simulate the optical, triggering, and
electronics response of the detector. The \textsc{SNOMAN} Monte Carlo
contains run-by-run detector state information, tracking
changes over time. All Monte Carlo was produced with at least
500 times the statistics expected in data.

For atmospheric neutrinos above 100~MeV,
we use \textsc{GENIE v2.12.2} \cite{Andreopoulos:2010eb,
Andreopoulos:2015vx} using the default model set, and the
Bartol04 flux predictions \cite{Barr:2004fw}, interpolated
between the solar minimum and maximum according to the dates
of each operational phase. The final state particles
from \textsc{GENIE} are then input into \textsc{SNOMAN} for propagation
through the full detector simulation.
Atmospheric neutrino oscillations are applied using
best-fit parameters in a model which
samples an ensemble of baselines from the neutrino
production height distributions of Gaisser and
Stanev \cite{Gaisser:1998bb}.

A model for final-state $\gamma$ resulting from interactions with oxygen is
included in the \textsc{GENIE} simulation
\cite{Andreopoulos:2015vx, Ejiri:1993ii, Kobayashi:2005ut}. However,
this does not include a potential background due to
a 15.1-MeV $\gamma$ produced in deexcitation of $^{12}$C$^\ast$
following neutrino interactions on $^{16}$O. Here, we take
a sample of such untagged $\gamma$ events following neutral current
quasielastic (NCQE)
interactions from a \textsc{NUANCE} (version 3r009) simulation,
which uses the calculation of Ejiri \cite{Ejiri:1993ii},
and scale according to the relative NCQE cross section
in \textsc{GENIE}. 

To model low-energy ($E_\nu<100$~MeV) atmospheric neutrino
interactions, we use the flux given by Battistoni {\it et al.}
\cite{2005APh....23..526B}; fluxes for the SNO location have been
provided by the authors. For this subdominant background, which
represents $\sim2$\% ($\sim4$\%) of the overall atmospheric neutrino
background in the $hep$ (DSNB) energy region of interest, only
$\nu_e$ and $\bar\nu_e$ are simulated, and the fluxes at the solar
minimum (when the background is largest) are used. This simulation is
performed directly in \textsc{SNOMAN}. We note that the low- and high-energy
atmospheric neutrino fluxes are the same as those used in the 2006
SNO $hep$ and DSNB search analysis \cite{Collaboration:2006wh}.

\subsection{Signals and backgrounds}
\label{sec:signals}
For the $hep$ solar neutrino signal, we use the spectrum computed
by Bahcall and Ulrich \cite{Bahcall:1987jc,Bahcall:1997wk}
and use the BSB05(GS98) flux of
$7.93(1\pm0.155)\times10^3$ cm$^{-2}$ s$^{-1}$
\cite{Bahcall:2006ke,Serenelli:2007zz}
as a benchmark.
The primary background for the $hep$ search is due to electrons
from $^8$B solar neutrino interactions, at a level that depends
on the shape of the spectrum near the end point. The spectral
shape from Winter {\it et al.} \cite{Winter:2006vi} is used, and
oscillations are applied according to a three-neutrino oscillation
model using best-fit parameters \cite{2010JPhG...37g5021N}. The
$^8$B solar neutrino flux is based on a three-phase analysis
of SNO $^8$B solar neutrino data, identical
to that presented in Ref. \cite{Aharmim:2013hr} except that
an upper energy threshold at 10~MeV was applied to eliminate any
contamination from a possible $hep$ signal. The extracted
$^8$B flux is
$\Phi_{{}^8\mathrm{B}} =
    (5.26
    \pm0.16\mathrm{~(stat.)}
    ^{+0.11}_{-0.13}\mathrm{~(syst.)})
    \times 10^6~\mathrm{cm}^{-2}~\mathrm{s}^{-1}$,
consistent with the published value.

The DSNB signal is modeled as an isotropic $\nu_e$ source using a
benchmark energy spectrum and total flux. We use the model of Beacom
and Strigari \cite{2006PhRvC..73c5807B} with $T=6$~MeV, which
predicts a total flux of
$\Phi^\mathrm{DSNB}_{\nu_e} = 0.66~\textrm{cm}^{-2}~\textrm{s}^{-1}$
in the energy range $22.9 < E_\nu < 36.9$~MeV.

Backgrounds due to isotropic light emission from the acrylic vessel
\cite{Aharmim:2005hw} have also
been studied using a dedicated event selection and Monte Carlo. The
background contamination depends on the choice of fiducial volume, and is
constrained to the negligible level of $<0.01$ events within our energy
regions of interest for the chosen cut of 550~cm.
Atmospheric neutrinos and associated $^{12}\mathrm{C}^*$ backgrounds
are modeled as described in \ref{sec:mc}. According to the \textsc{GENIE} simulation,
the dominant source of atmospheric background is from decay at rest of
muons below or near the Cherenkov threshold. These are predominantly
produced directly in $\nu_\mu$ and $\bar\nu_\mu$ CC interactions, with
a small contribution from decays of subthreshold CC- and NC-produced
$\pi^\pm\to\mu^\pm\to e^\pm$. Decays of subthreshold muons account for
the majority of the background for the DSNB search, while the atmospheric
backgrounds for the $hep$ search are subdominant and result from a mix of
subthreshold muon decays, 15.1-MeV $\gamma$ rays, and other NC interactions.
The direct production of untagged low-energy electrons in $\nu_e$ CC
interactions accounts for a small portion of the background, $\lesssim10$\%
in each case.

\begin{figure}
\centering
\includegraphics[width=0.5\textwidth]{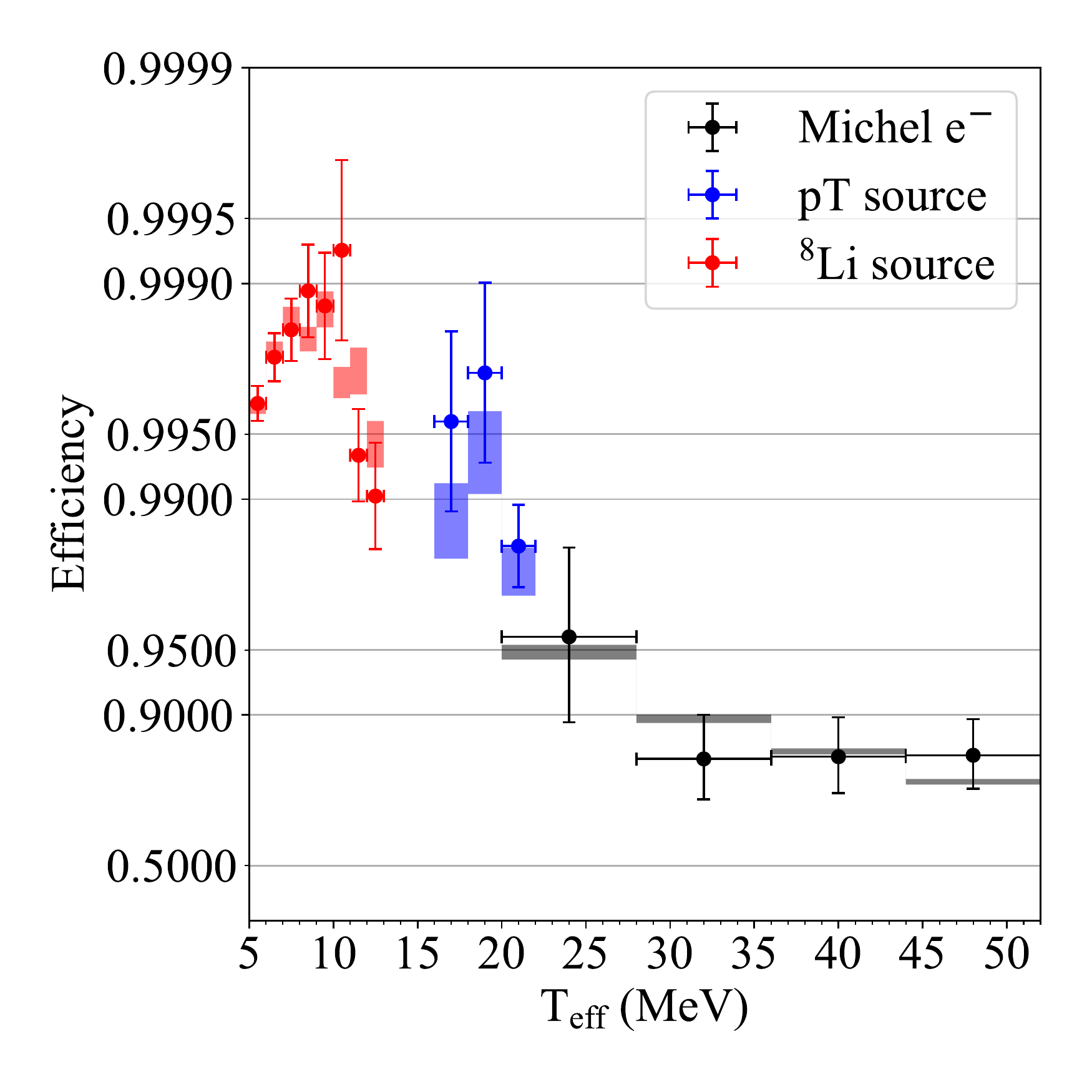}
\caption{Efficiency of the high-level event selection cuts for phase I,
compared between calibration sample data (points) and Monte Carlo (shaded
boxes). The calibration samples include deployed
$^{8}$Li \cite{Tagg:2002iq} and $pT$ \cite{Poon:403099} sources
and Michel electrons from muons that stop and decay inside the detector.}
\label{fig:hlc-sacrifice-d2o}
\end{figure}

\subsection{Counting analysis}
\label{sec:counting}
Within each energy region of interest (ROI) for the single-bin counting
analysis, 1D cuts on high level features are simultaneously tuned to optimize
the search sensitivity in Monte Carlo, with further adjustments to
minimize the impact of the systematic uncertainties on the shapes of the
observable distributions. The $hep$ energy ROI of $14.3<T_\mathrm{eff}<20$~MeV
and DSNB ROI of $20<T_\mathrm{eff}<40$~MeV are chosen to optimize
signal-to-background ratio while maximizing signal acceptance, following the
procedure described in Ref. \cite{Collaboration:2006wh}.
The signal efficiency of the high-level cuts is validated using calibration
datasets as shown in Fig. \ref{fig:hlc-sacrifice-d2o}.
Within the $hep$ ROI, the high level and burst cuts together reduce the
atmospheric neutrino backgrounds by 97\%, with a signal efficiency
of $\sim99$\%.

For the purposes of this cut-based analysis,
confidence intervals are constructed using a Bayesian framework in which we
construct intervals from a Poisson likelihood function marginalized over the
expected background distribution. This function is defined as
\begin{equation}
\begin{split}
-\log \mathcal{L}(\mu, b | n, \hat{b}, \sigma_b) = \mu &+ b \\
   &+ \log \Gamma(n + 1) \\
   &- n \times \log(\mu + b) \\
   &+ \left. \frac{1}{2}\frac{(b - \hat{b})^2}{\sigma_b^2} \right.,
\end{split}
\end{equation}
where $\mu$ is the true signal mean, $b$ the true background rate, $n$ the
observed number of events, $\hat{b}$ the mean background expectation, and
$\sigma_b$ the Gaussian uncertainty on $b$.
In constructing this likelihood function, we have chosen a step function prior
that is constant for $\mu>0$.
Integrating over the background parameter $b$ yields
$-\log \mathcal{L}(\mu|n)$, which is treated as a posterior probability
distribution function (PDF) for $\mu$ and
used to construct intervals.
For a confidence level $\alpha$, a
two-sided interval $\mathcal{C}$ is defined by the highest posterior density
region (HPDR), i.e. adding points $\mu$ in order of their posterior probability
density until $\sum_\mathcal{C}\mathcal{L}(\mu|n)\geq\alpha$. One-sided
intervals are constructed by direct integration of $\mathcal{L}$ to determine
the smallest
$\mu'$ such that $\sum_0^{\mu'} \mathcal{L}(\mu|n) \geq \alpha$.

\subsection{Likelihood analysis}
\label{sec:sigex}
In order to leverage the energy dependence of the signal spectra and
lower the threshold for the $hep$ search, an unbinned
maximum likelihood fit was
also performed. The fit considers all three phases simultaneously,
with the $^8$B and $hep$ fluxes held constant across time, as well as the
overall atmospheric neutrino flux normalization after
accounting for differences across the solar minimum and maximum.
The dominant systematic uncertainties are varied in
the fit using Gaussian pull terms, and include the oscillation parameters
$\theta_{12}$ and $\Delta m^2_{12}$ as well as the energy scale and
resolution model parameters and angular and $\beta_{14}$ resolutions, which are
treated as uncorrelated. The fit uses three-dimensional PDFs, binned in
reconstructed energy ($T_\mathrm{eff}$, ten bins, $10-20$~MeV),
the angle relative to the Sun ($\cos\theta_\mathrm{sun}$, ten bins, $-1-1$),
and the isotropy parameter ($\beta_{14}$, 15 bins, $-0.12-0.95$).
PDFs are constructed for $^8$B CC electrons, $^8$B ES electrons,
$hep$ CC electrons, $hep$ ES electrons, and atmospheric neutrino interactions
for each phase. The relative
normalizations of the CC and ES components for each signal are
fixed. The cuts described previously are applied to
data and Monte Carlo prior to PDF construction and fitting; these
include the fiducial volume, ITR, three KS
probability figures of merit, and low-level cuts. In contrast to
the counting analysis, energy and isotropy are observables in the
fit.

The full negative log likelihood function optimized in the fit is of the form
\cite{Seibert:2008vh}
\begin{equation}
\label{eq:ml-full}
\begin{split}
-\log &\mathcal{L}({\bf r},\Delta) = \\
& \sum_{j=1}^M \tilde{N}_j({\bf r}, \Delta) \\
-&\sum_{i=1}^N \log \left( \sum_{j=1}^M \tilde{N}_j({\bf r}, \Delta) \times \mathbf{P}_j({\bf x}_i, \Delta) \right) \\
+&\frac{1}{2} \sum_{k=1}^{M'} \frac{(r_k - \bar{r}_k)^2}{\sigma_{r_k}^2} \\
+&\frac{1}{2} \sum_{m=1}^{s} \frac{(\Delta_m - \bar{\Delta}_m)^2}{\sigma_{\Delta_m}^2},
\end{split}
\end{equation}
where the first term corresponds to the total normalization constraint, the
second to the unbinned likelihood given the PDFs, and the final two terms
represent Gaussian uncertainties on rate and systematic parameters.
In Eq. \ref{eq:ml-full},
$\mathbf{P}_j$ are PDFs for each signal, which are binned in the set of observables
$\mathbf{x}$. These PDFs
are constructed from Monte Carlo events that have been modified
according to a set of $s$ systematic parameters $\Delta$, with
associated Gaussian uncertainties $\sigma_\Delta$.

The parameters ${\bf r}$ correspond to signal rates, which may be correlated
across event types, e.g. in the case of the $hep$ flux which scales both
the CC and ES $hep$ event rates. Thus, the signal
rates are related to the expected number of events of a particular type
($\tilde{N}$) by an efficiency matrix $\epsilon$ defined such
that $\tilde{N}_i = \epsilon\indices{_i^j}r_j$. The Gaussian
uncertainties associated with signal rates are denoted $\sigma_r$.
$M'$ is simply the number of rate parameters which are externally constrained.

The efficiency matrix $\epsilon$ also accounts for events shifting into
or out of the boundary of the analysis window (a volume $V$ in observable
space) following the application of a systematic transformation $S$. This
is handled by the inclusion of a weighting factor
$|\{\mathbf{x}_i | S(\mathbf{x}_i, \Delta) \in V \}|$.

The fit was performed using a Markov chain Monte Carlo (MCMC), and
a number of metrics were used to evaluate fit quality and convergence. These
included a
check of statistical compatibility of parameter distributions within
subdivisions of the MCMC random walk, and a toy Monte Carlo to evaluate the
goodness of fit through a $\chi^2$ hypothesis test. Additional validation
included signal injection tests varying the $hep$ flux from $0.01-10$ times
model predictions.

\subsection{Systematic uncertainties}
\label{sec:syst}
A number of systematic effects are important within these analyses.
The primary background to the $hep$ search is electrons from
$^8$B solar neutrino interactions, where the spectrum is affected by the energy
response modeling as well as the flux normalization and intrinsic
shape. The flux uncertainty is taken from the three-phase fit
to low-energy SNO data described in Sec. \ref{sec:signals}, and
the shape is varied within the uncertainties provided by
Winter {\it et al.} \cite{Winter:2006vi}. For solar neutrinos,
the uncertainties in the oscillation parameters and
the $\nu-d$ CC cross section are also included.
To address the energy response modeling, which affects all
signals and backgrounds, uncertainties are derived from fits
to deployed calibration sources and samples of Michel
electrons; this procedure is described in Sec.
\ref{sec:response}. Uncertainties impacting atmospheric
neutrino backgrounds are detailed in Sec. \ref{sec:atm-syst}.
The major systematic uncertainties impacting the
analyses are summarized in Table \ref{tab:systs}.
\begin{table}
\begin{ruledtabular}
\caption{Systematic uncertainties. Values apply to all three phases except
as noted for those in the lower part of the table.}
\label{tab:systs}
\centering
\begin{tabular}{lccc}
Parameter                    && Magnitude &\\
\hline
Vertex accuracy              && 2.9\% \cite{Aharmim:2013hr} &\\
Vertex resolution            && 2.4~cm \cite{Aharmim:2013hr} &\\
Angular resolution           && 2\%      &\\
$^8$B flux                   && See Sec. \ref{sec:signals}    & \\
$^8$B $\nu_e$ spectrum       && Ref. \cite{Winter:2006vi}       &\\
$\nu$ Mixing parameters      && Ref. \cite{2010JPhG...37g5021N} &\\
Atmospheric $\nu$ flux&&&\\
~~~$E_\nu > 100$~MeV         && 10\% \cite{Barr:2004fw}   &\\
~~~$E_\nu < 100$~MeV         && 25\% \cite{2005APh....23..526B}    &\\
Cross sections&&&\\
~~~CC $\nu-D$                && 1.2\%    &\\
~~~Atmospheric $\nu$                && See Sec. \ref{sec:atm-syst} &\\
~~~15.1~MeV $\gamma$ rays    && 100\%    &\\
&&&\\
\hline
                             & Phase I & Phase II & Phase III \\
\hline
Live time                    & 0.006\%       & 0.021\%        & 0.36\% \\
Energy scale&&&\\
~~~$T_\mathrm{eff}=14.3$~MeV & 0.61\%        & 0.55\%         & 0.82\% \\
~~~$T_\mathrm{eff}=20.0$~MeV & 0.71\%        & 0.65\%         & 0.86\% \\
Energy resolution scale      & 1.60\%        & 1.71\%         & 1.37\% \\
\end{tabular}
\end{ruledtabular}
\end{table}

\subsubsection{Detector response}
\label{sec:response}
\begin{table*}
\begin{ruledtabular}
\caption{Data/Monte Carlo comparisons for number of followers in selected
atmospheric neutrino event candidate events. Followers with
$\Delta t <(>) 20~\mu$s are primarily Michel electrons (neutrons).}
\label{tab:followers}
\centering
\begin{tabular}{lcccccc}
& \multicolumn{2}{c}{Phase I} & \multicolumn{2}{c}{Phase II} & \multicolumn{2}{c}{Phase III} \\
& Data & MC & Data & MC & Data & MC \\
\hline
All followers      & 59 & $59.19\pm11.52$ & 184 & $180.77\pm26.70$ & 72 & $62.04\pm10.30$ \\
$\Delta t<20~\mu$s & 25 & $30.42\pm 7.06$ &  31 & $ 48.09\pm 9.23$ & 39 & $39.25\pm 7.49$ \\
$\Delta t>20~\mu$s & 34 & $28.77\pm 6.89$ & 153 & $132.68\pm20.90$ & 33 & $22.80\pm 5.56$ \\
\end{tabular}
\end{ruledtabular}
\end{table*}
In order to calibrate the response in the detector across an
energy range up to 40~MeV, several event samples were compared against
\textsc{SNOMAN} Monte Carlo predictions. The vertex reconstruction is
described in Refs.
\cite{Aharmim:2007hh} (phases I and II) and
\cite{Aharmim:2013gz} (phase III), and based on this we include a 2.4~cm
uncertainty on reconstructed position resolution, and an overall 2.9\%
fiducial volume uncertainty.
Additionally, a 2\% uncertainty on the angular resolution
for ES events is modeled as a scaling via a parameter $\Delta_\theta$
\cite{Aharmim:2013hr}:
\begin{equation}
(\cos\theta)' = 1 + (\cos\theta - 1) (1 + \Delta_\theta),
\end{equation}
where $(\cos\theta)'$ outside the interval $[-1,1]$ are assigned a random value
within that interval.

In each of the three phases, a large sample of 6.13-MeV $\gamma$ rays from a
deployed $^{16}$N source provided the primary calibration. Additionally,
a $pT$ source
in phase I provided a sample of 19.8-MeV $\gamma$ rays \cite{Poon:403099}.
To extend the model to higher energies, samples of Michel electrons
from decays of stopping cosmic ray muons
were selected
and fit to a response model allowing an energy-dependent fractional
energy scaling ($\Delta_S^{(i)}$) and shift in resolution ($\Delta R$):
\begin{equation}
\label{eq:escale-linear}
T_\mathrm{eff}' = T_\mathrm{eff} +
  (\Delta_S^{(0)} + \Delta_S^{(1)}\cdot T_\mathrm{eff}) \cdot T_\mathrm{eff} +
  \Delta_R \cdot (T_\mathrm{eff} - T_\mathrm{true}).
\end{equation}
The parameters were extracted using a maximum likelihood fit to the
Michel electron samples for each phase, subject to prior constraints
based on the deployed source measurements. The extracted parameters
are given in Table \ref{tab:resp-params}. We find that the parameters are
consistent with zero, confirming that the initial $^{16}$N-based energy
calibration provides a reasonable estimate of energy across the regions of
interest, and the correlated errors in each phase indicate the magnitude of
systematic shifts that remain compatible with the higher-energy calibration
samples. This provides a data-driven constraint on the smearing of the
spectrum of electrons produced by $^{8}$B solar neutrino interactions, which
forms a dominant background for the $hep$ search.

\begin{table}
\begin{ruledtabular}
\caption{Energy response model parameters extracted from maximum likelihood
fits to calibration sample data in each phase.}
\label{tab:resp-params}
\centering
\begin{tabular}{lccc}
Parameter & Phase I & Phase II & Phase III \\
\hline
Normalization &
$135\pm12.2$   & $213\pm14.8$   & $172\pm13.0$  \\
$\Delta_S^{(0)}/10^{-3}$ &
$-5.20\pm7.21$ & $-0.01\pm6.14$ & $ 1.25\pm10.2$ \\
$\Delta_S^{(1)}/10^{-3}$ &
$ 0.44\pm0.42$ & $-0.16\pm0.37$ & $-0.16\pm0.43$ \\
$\Delta_R^{(0)}/10^{-2}$ &
$ 1.83\pm1.60$ & $ 2.38\pm1.71$ & $ 1.61\pm1.37$ \\
\end{tabular}
\end{ruledtabular}
\end{table}

Additionally, a similar model including a linear scaling and resolution was
applied to the shape of the isotropy parameter $\beta_{14}$, with
$\Delta_S^{(0)} = \Delta_R = 4.2\times10^{-3}$ for all three phases, based
on measurements with the $^{16}$N calibration source \cite{Aharmim:2013hr}. 
Finally, the contribution of any non-Gaussian (flat) tails
in the energy response was constrained to the level of
$\lesssim10^{-3}$ events in
the energy region of interest based on samples of events from the
deployed $^8$Li source \cite{Tagg:2002iq},
which has a $\beta$ spectrum similar to that of the $^8$B solar neutrinos.

\subsubsection{Atmospheric neutrinos}
\label{sec:atm-syst}
Two main classes of uncertainty affect the atmospheric neutrinos:
the flux uncertainty, which is
taken to be 25\% \cite{2005APh....23..526B} and
10\% \cite{Barr:2004fw} for low ($<0.1$~GeV) and high ($0.1-10$~GeV)
energies, respectively,
and the cross sections. The cross section uncertainties are
evaluated through event reweighting, by simultaneously varying
the parameters in the default \textsc{GENIE} model set (see Ref.
\cite{Andreopoulos:2015vx}) within their
respective uncertainties to produce an ensemble of weights
corresponding to different model hypotheses.

To validate the modeling of atmospheric neutrino interactions, a
sample of fully-contained atmospheric neutrino events was selected.
These events are required to have $200-5000$ PMTs hit, no activity in the
veto region, and must not follow an event tagged with a $\mu$ entering the
detector. These requirements provide a high-purity sample of contained
atmospheric neutrino candidate events with $T_\mathrm{eff}\geq25$~MeV that
is independent from the signal selection.
Starting from this selection, we search for time-coincident follower events,
which mainly consist of Michel electrons ($\Delta t<20~\mu$s) and neutrons
($\Delta t>20~\mu$s). These follower events must pass all analysis
cuts and have an energy $5<T_\mathrm{eff}<100$~MeV. For the
selected events, we compare the multiplicity and timing of coincidences
as well as the energy, position, isotropy, and other high-level observables
between the atmospheric Monte Carlo and data, and find good agreement within
the flux and cross section modeling uncertainties of the \textsc{GENIE} Monte Carlo
simulation. Table \ref{tab:followers}
provides the total number of atmospheric neutrino candidate follower
events, compared to the Monte Carlo expectation. 

Additionally, a search was performed for events in the energy range
$35<T_\mathrm{eff}<70$ MeV, where Michel electrons from atmospheric
neutrinos are expected. With all event selection cuts applied, six
isolated events were observed, with a Monte Carlo expectation of 3.7 ($p$-value
17\%). The event rates are consistent across phases, with one event observed (0.8
expected) in phase I, three observed (1.3 expected) in phase II, and two observed
(1.6 expected) in phase III. Relaxing the time coincidence cuts, we find one,
two, and three additional events in phases I, II, and III, respectively. 
Of these six events, one is followed by a neutron candidate event. The other
five are preceded within a few $\mu$s by a low-energy event,
of which three are consistent with deexcitation photons from the
primary neutrino interaction, and two are most likely to be
near-threshold atmospheric neutrino-induced muons. Extending to higher
energies, $70<T_\mathrm{eff}<100$ MeV, we find one additional electronlike
event in phase II, which appears in isolation.
Further details of these events may be found
in Table \ref{tab:he-events} in the Appendix.

\section{Results}
\label{sec:results}
\subsection{Counting analysis}
\begin{figure*}
\centering
\includegraphics[width=\textwidth]{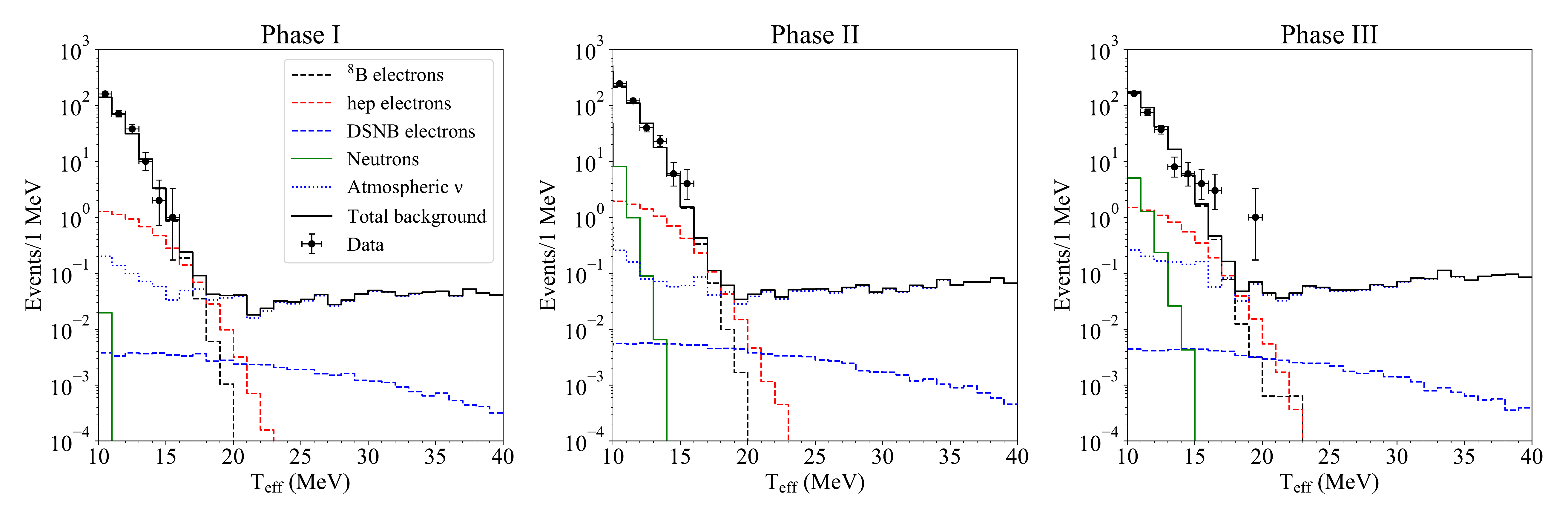}
\caption{Reconstructed energy spectra for each phase.}
\label{fig:counting-spectra}
\end{figure*}

Within the sensitivity-optimized energy regions of interest for the $hep$ and
DSNB signals, we performed a single-bin counting analysis as introduced in
Sec. \ref{sec:counting}.
The energy spectra for selected events are shown in Fig.
\ref{fig:counting-spectra}.
The total signal and background expectations in
the $14.3-20$~MeV $hep$ energy ROI are $3.09\pm0.12$ and $13.89\pm1.09$,
respectively, with 22 events observed. Nearly all the background in the $hep$
ROI is due to $^{8}$B solar neutrinos. In the DSNB ROI, 0.08 signal events
and 2.58 background events are expected, with zero events observed.
The distribution across phases is
given in Table \ref{tab:counting-counts}.

The uncertainties on the total three-phase
signal and background expectations are correlated
($r_{hep}=0.83$, $r_\mathrm{DSNB}=0.12$), and are obtained
using an ensemble of 500 three-phase pseudo-experiments with systematic
parameters randomly sampled according to their correlated uncertainties.
The dominant source of uncertainty in the $hep$ region is the energy
response modeling, due to the steeply falling tail of the $^8$B solar neutrino
backgrounds. This model is constrained using data spanning the energy
range as described in Sec. \ref{sec:response}.

The majority of candidate events, 13 of 22, occurred during
phase III. These events appear signal-like in all respects, and consistency
with background is observed in sidebands with respect to energy and all other
high-level observables. According to toy Monte Carlo studies,
the probability of observing a statistical fluctuation of at least
this magnitude in any one phase is approximately 8\%.

Applying the Bayesian procedure described in Sec.
\ref{sec:counting} yields an 68.3\% credible interval (CI) of 
$\Phi_{hep} = (9.6 - 33)\times10^{3}~\mathrm{cm}^{-2}~\mathrm{s}^{-1}$;
however, as the probability of a statistical fluctuation of this magnitude is
significant, we set a one-sided upper limit of
\[
\Phi_{hep} < 40\times10^{3}~\mathrm{cm}^{-2}~\mathrm{s}^{-1}
~(90\%~\mathrm{CI}).
\]
For comparison, in the previous phase I analysis two events were observed
with $0.99\pm0.09$ signal and $3.13\pm0.60$ background events
expected; this resulted in a 90\% CL frequentist upper limit on the $hep$
flux of $23\times10^{3}~\textrm{cm}^{-2}~\textrm{s}^{-1}$ \cite{Collaboration:2006wh}.

Of the 2.58 events expected in the DSNB ROI, 2.47 are due to high-energy
($E_\nu>100$~MeV) atmospheric neutrinos. 82\% of these are CC interactions,
of which 80\% are due to the decay of muons below the Cherenkov threshold,
and in 10\% an isolated electron is directly produced in a $\nu_e$ CC
interaction.
Of the 18\% NC contribution, $\sim75$\% are due to subthreshold
muon decays following charged meson production.
The remaining 0.11 expected events are due to low-energy
($E_\nu<100$~MeV) atmospheric neutrinos, with about 90\% $\nu_e$ and
10\% $\bar\nu_e$.
The median experiment in a Monte Carlo ensemble provides 90\% CI
sensitivity to signals at least 52 times larger than the benchmark Beacom
and Strigari $T=6$~MeV model.
With an apparent downward fluctuation, zero events are observed, and
we set an upper limit of 29 times the model prediction,
corresponding to DSNB $\nu_e$ flux of
$\Phi^\mathrm{DSNB}_{\nu_e} < 19~\textrm{cm}^{-2}~\textrm{s}^{-1}$
(90\% CI) in the energy range $22.9 < E_\nu < 36.9$~MeV. The dominant
source of systematic uncertainty in the DSNB ROI is the 10\% normalization
uncertainty for the flux of atmospheric neutrinos with $E_\nu>100$~MeV.

\begin{table}
\begin{ruledtabular}
\caption{Summary of expected and observed events for each
ROI and phase in the counting analysis.}
\label{tab:counting-counts}
\centering
\begin{tabular}{lccc}
& Expected & Expected   & Events \\
& signal   & background & observed \\
\hline
Phase I $hep$   & $0.84 \pm 0.08$ & $ 3.14 \pm  0.63$ &  3 \\
Phase II $hep$  & $1.28 \pm 0.06$ & $ 5.37 \pm  0.65$ &  6 \\
Phase III $hep$ & $0.98 \pm 0.05$ & $ 5.38 \pm  0.52$ & 13 \\
Total $hep$     & $3.09 \pm 0.12$ & $13.89 \pm  1.09$ & 22 \\
\hline
Phase I DSNB    & $0.02 \pm 0.00$ & $ 0.62 \pm  0.10$ &  0 \\
Phase II DSNB   & $0.03 \pm 0.00$ & $ 0.91 \pm  0.15$ &  0 \\
Phase III DSNB  & $0.02 \pm 0.00$ & $ 1.06 \pm  0.17$ &  0 \\
Total DSNB      & $0.08 \pm 0.00$ & $ 2.58 \pm  0.26$ &  0 \\
\end{tabular}
\end{ruledtabular}
\end{table}

\subsection{Likelihood analysis}
\label{sec:sigex-results}
\begin{figure*}
\centering
\includegraphics[width=\textwidth]{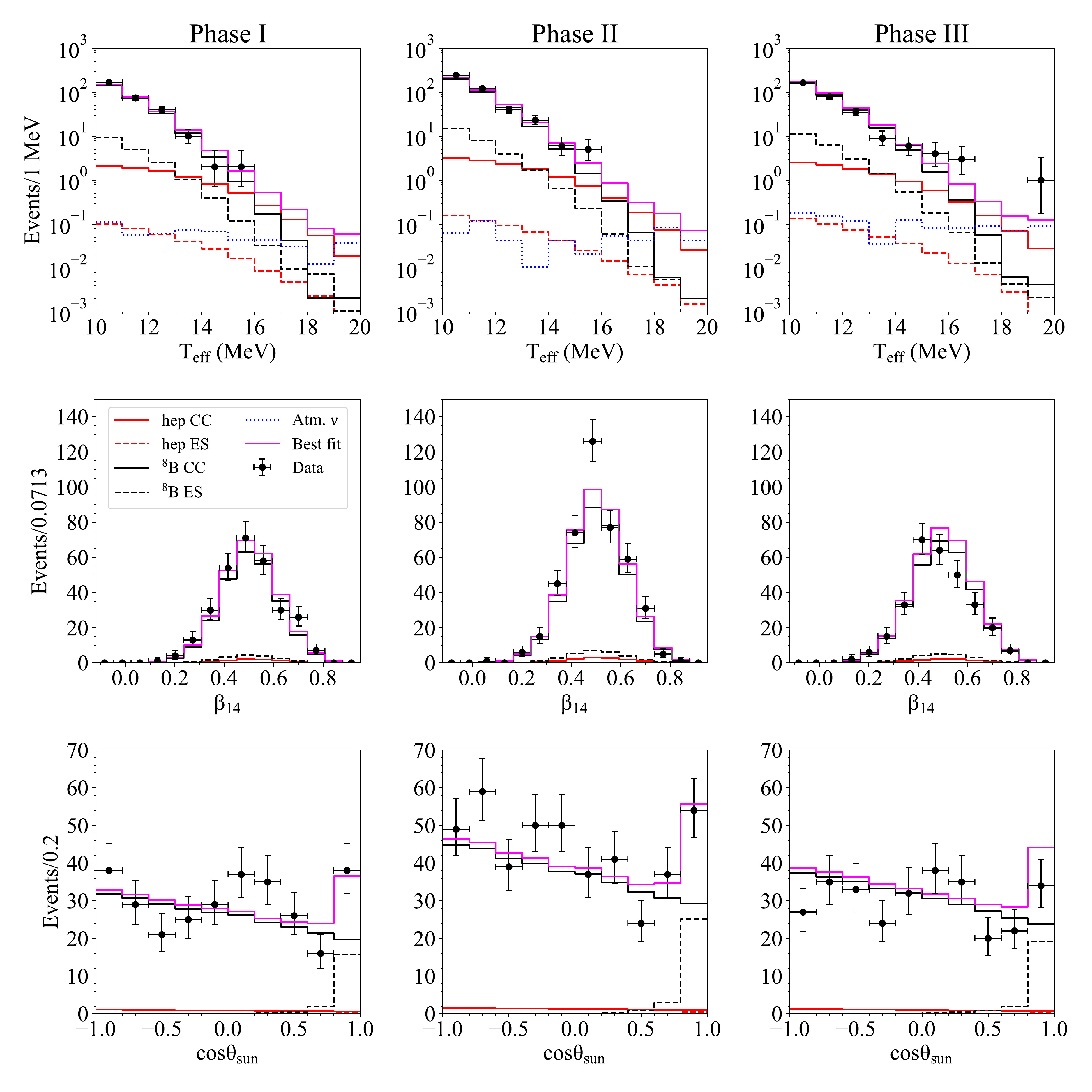}
\caption{Distributions of events in the full dataset compared to the
best fit in the joint three-phase likelihood analysis, with projections shown
for each phase and fit observable. Distributions are shown over the full
energy range of the fit, $10-20$~MeV. The model and systematic uncertainties
are discussed in Secs. \ref{sec:sigex} and \ref{sec:syst}, respectively,
with the extraction of the $hep$ flux described in Sec.
\ref{sec:sigex-results}.}
\label{fig:sigex-all-fits}
\end{figure*}

For the $hep$ search, we additionally performed a likelihood fit as
described in Sec. \ref{sec:sigex}.
One-dimensional projections of the best fit in the observable dimensions
$T_\mathrm{eff}$, $\beta_{14}$, and $\cos\theta_\mathrm{sun}$
are shown in Fig.
\ref{fig:sigex-all-fits}. We note that the shape of the
$\cos\theta_\mathrm{sun}$ is determined by the $\nu_e$ ES and CC cross sections;
in the former the outgoing electron direction is strongly correlated with the
incoming neutrino direction, while in the latter it is moderately
anticorrelated.
The quality of the fit was evaluated using a $\chi^2$ test based on
an effective test statistic distribution derived using a toy Monte Carlo,
yielding a $p$ value of 16.0\% considering statistical errors only.

\begin{figure}
\centering
\includegraphics[width=0.5\textwidth]{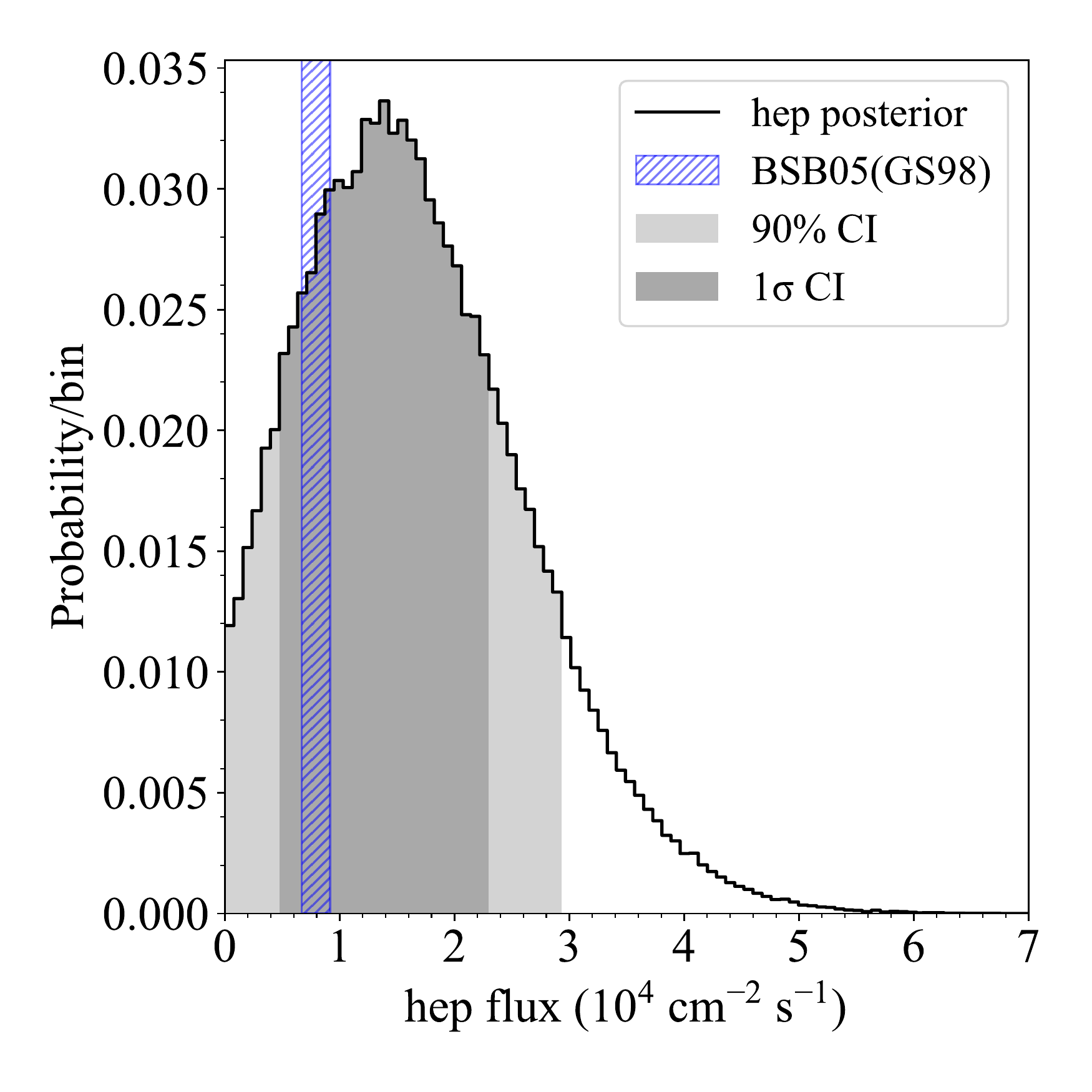}
\caption{The posterior distribution for the $hep$ flux, marginalized over all
other fit parameters, with the 90\% and $1\sigma$ credible intervals. The
BSB05(GS98) standard solar model prediction
\cite{Bahcall:2006ke,Serenelli:2007zz} is also shown for comparison.}
\label{fig:hep-posterior}
\end{figure}
Bayesian credible intervals are obtained as within the counting analysis, by
marginalizing over all other parameters. The $1\sigma$ and 90\% credible intervals
are shown in Fig. \ref{fig:hep-posterior}. We note that the intervals and
best-fit value obtained with this Bayesian approach are consistent with
quantities obtained by directly analyzing the likelihood space sampled by the
MCMC.

In agreement with the counting analysis up to differences introduced
by the statistical treatments, this result is compatible with the BSB05(GS98)
model prediction and is consistent with zero $hep$ flux. 
The fit yields a $68.3\%$ HPDR credible interval for the $hep$ flux parameter
corresponding to
$\Phi_{hep} = (5.1-23)\times10^{3}~\mathrm{cm}^{-2}~\mathrm{s}^{-1}$;
as in the counting-based analysis, we define a one-sided upper limit:
\[
\Phi_{hep} < 30\times10^{3}~\mathrm{cm}^{-2}~\mathrm{s}^{-1}
~(90\%~\mathrm{CI}).
\]

\section{Conclusions}
\label{sec:conclusion}
Data from the full SNO dataset, representing an exposure of 2.47 kilotonne
years
with a D$_2$O target, has been analyzed to search for neutrinos from the
$hep$ reaction in the Sun's $pp$ chain and $\nu_e$ from the diffuse supernova
neutrino background. In addition to increasing the exposure by a factor of
3.8 relative to the previous SNO search for these signals
\cite{Collaboration:2006wh}, a new spectral fit has been employed to improve
the sensitivity to the $hep$ flux.

We have performed the most sensitive search to date for the $hep$ solar
neutrino flux, the final unobserved branch of the $pp$ fusion chain.
This measurement is compatible with the BSB05(GS98) model prediction
of $(7.93\pm1.23)\times10^{3}~\mathrm{cm}^{-2}~\mathrm{s}^{-1}$,
while remaining consistent with zero $hep$ flux, and we extract
a one-sided upper limit of
$\Phi_{hep} < 30\times10^{3}~\mathrm{cm}^{-2}~\mathrm{s}^{-1}~90\%~\mathrm{CI}$.
In a search at energies above the solar neutrino end points,
we observe no evidence for the DSNB $\nu_e$ flux, and set an upper limit on
this flux; our results suggest that a $\nu_e$ flux larger than $\sim30$ times
the current predictions is disfavored.
Upcoming experiments sensitive to DSNB $\bar\nu_e$ through
inverse beta decay anticipate sensitivity at the level of model predictions
\cite{PhysRevLett.93.171101,MartiMagro:2018ksp,An:2015jdp}.
Additionally, the DUNE experiment \cite{PhysRevLett.123.131803, DUNETDR} and
other future large detectors may offer improved sensitivity to both $hep$ and
DSNB neutrinos.

\begin{acknowledgments}
This research was supported by: Canada: Natural Sciences and Engineering
Research Council, Industry Canada, National Research Council, Northern Ontario
Heritage Fund, Atomic Energy of Canada, Ltd., Ontario Power Generation, High
Performance Computing Virtual Laboratory, Canada Foundation for Innovation,
Canada Research Chairs program, Breakthrough Prize Fund at Queen’s University;
U.S.: Department of Energy Office of Nuclear
Physics, National Energy Research Scientific Computing Center, Alfred P. Sloan
Foundation, National Science Foundation,
Department of Energy National Nuclear Security Administration through the
Nuclear Science and Security Consortium; U.K.: Science and Technology Facilities
Council (formerly Particle Physics and Astronomy Research Council); Portugal:
Funda\c{c}\~{a}o para a Ci\^{e}ncia e a Tecnologia. We thank the SNO technical
staff for their strong contributions. We thank INCO (now Vale, Ltd.) for
hosting this project in their Creighton mine. We also thank John Beacom for
helpful suggestions and discussions regarding atmospheric neutrino backgrounds.
\end{acknowledgments}

\appendix
\section*{Appendix: High-Energy Sideband}
\label{sec:he-events}
Table \ref{tab:he-events} provides details on events discussed in
Sec. \ref{sec:atm-syst}. These events are selected at energies
$35<T_\mathrm{eff}<100$ MeV.

\begin{table*}
\begin{ruledtabular}
\caption{Details of selected high-energy sideband events. Where associated
time-correlated events are present, the time difference $\Delta t$ relative
to the selected electron-like event is given, along with the reconstructed
energy of the coincidence event, $E_\mathrm{coinc}$.}
\label{tab:he-events}
\centering
\begin{tabular}{ccccccc}
Phase & $N_\mathrm{hits}$ & $E$ (MeV) & Radius (cm) & Coincidence & $\Delta t$ & $E_\mathrm{coinc}$ \\
\hline
I   & 379 & 48.1 & 247.9 & $\gamma$ candidate & $-1.2$~$\mu$s &  3.9 MeV \\
I   & 452 & 56.8 & 143.5 &     &&\\
\hline
II  & 330 & 42.0 & 432.6 &      &&\\
II  & 369 & 44.5 & 320.4 & $\mu$ candidate    & $-0.8$~$\mu$s & 16.3 MeV \\
II  & 401 & 58.3 & 104.0 & $\gamma$ candidate & $-1.7$~$\mu$s &  4.3 MeV \\
II  & 472 & 67.5 & 173.2 &      &&\\
II  & 400 & 57.6 & 487.7 &     &&\\
II  & 633 & 81.9 & 405.9 &     &&\\
\hline
III & 313 & 43.0 & 386.6 & $\gamma$ candidate & $-1.3$~$\mu$s &  2.4 MeV \\
III & 348 & 47.5 & 179.4 & $\mu$ candidate    & $-0.6$~$\mu$s &  9.4 MeV \\
III & 265 & 38.2 & 354.4 &     &&\\
III & 258 & 36.0 & 539.8 & $n$ candidate      &      $+18$ ms &  5.5 MeV \\
III & 326 & 47.3 & 487.7 &     &&\\
\end{tabular}
\end{ruledtabular}
\end{table*}

\bibliography{bibliography}{}
\bibliographystyle{apsrev4-1}
\end{document}